\documentstyle[aps,prl,multicol,epsfig]{revtex}
\newcommand{\beq}{\begin{equation}}
\newcommand{\eeq}{\end{equation}}
\newcommand{\beqa}{\begin{eqnarray}}
\newcommand{\eeqa}{\end{eqnarray}}
\newcommand{\ax}{a^{\dagger}}
\newcommand{\cx}{c^{\dagger}}

\newcommand{\dinf}{$z\rightarrow\infty$ }

\begin{document}
\title{Optical absorption from a  non degenerate polaron gas}

\author{S. Fratini and F. de Pasquale}

\address{Istituto Nazionale di Fisica della
Materia and Dipartimento di Fisica\\
Universit\`a di Roma "La Sapienza",
Piazzale Aldo Moro 5, I-00185 Roma, Italy}

\author{S. Ciuchi} 

\address{Istituto Nazionale di Fisica della Materia and 
Dipartimento di Fisica\\
Universit\`a dell'Aquila,
via Vetoio, I-67010 Coppito-L'Aquila, Italy}

\date{\today}
\maketitle

\begin{abstract}
We apply the Dynamical Mean Field Theory to calculate the low-density
limit of the optical conductivity in the Holstein model.
This non perturbative treatment
allows to span continuously from 
the weak-coupling quasi-free electron behaviour
to the strong-coupling small polaron picture. New features arise
in the intermediate coupling regime:
i) a noticeable transfer of spectral weight towards low frequencies occurs at
low temperatures, when coherent processes acquire importance; 
ii) thermally activated Polaron Interband Transitions appear 
at non zero temperature, whose spectral weight has a
non monotonic temperature behaviour.
We propose an interpretation of the infrared active vibrations (IRAV)
observed in the optical
spectra of the underdoped superconducting cuprates in terms of small polarons 
in the intermediate coupling regime.
\end{abstract}
\begin{multicols}{2}    

The calculation of the optical conductivity of small polarons is an 
important and unsolved problem in solid state physics. Although many
approaches exist which are appropriate in specific limiting cases,
an analytical treatment which is able to give the correct
physical results in all the regimes of parameters is still lacking.
The simplest model which describes small polarons is 
the Holstein Molecular Crystal Model \cite{Holstein}, 
where the carriers interact with
phonon modes localized on each lattice cell. 
The Holstein model 
  was recently solved \cite{PRB} in the single-electron case by  Dynamical
  Mean Field Theory (DMFT) \cite{Kotliar}, which becomes
  exact in the limit of infinite lattice coordination. 
  When applied to real lattices, this approach can be viewed as a
  valid interpolation scheme, provided that an appropriate finite dimensional density of
  states (DOS) is chosen \cite{Piovra,Sumi}. The solution reported in ref. 
  \cite{PRB} is essentially  analytical, and it overcomes the well-known
  difficulties inherent to numerical methods. 

The aim of this work is
  to generalize the exact DMFT solution of the
  Holstein model to the calculation of the optical conductivity. 
We shall focus
on the most interesting \textit{intermediate coupling
regime} near the adiabatic limit, 
where novel features have been found in the single
particle propagator, so that new interesting properties can be
expected in the optical spectra.
Contrary to the usual 
strong coupling situation \cite{Reik}, where the polaron absorption is
concentrated in a ``band'' located at frequencies much higher than the phonon
frequency, we will
show that at intermediate coupling strengths, narrow absorption peaks appear
close to the phonon frequencies, which are due to transitions between
different subbands in the polaron excitation spectrum. The presence of
such low-frequency features is a very general
property of finite band electron-phonon (e-ph) systems in the intermediate
coupling regime, which could help interpreting the far infra-red
spectra of the superconducting cuprates at low doping levels, 
as discussed in the conclusion.

  In the Holstein hamiltonian \cite{Holstein}, tight-binding electrons
  with hopping amplitude $t$ are coupled locally to Einstein phonons
  with energy $\omega_0$\cite{Footnote2}:
\begin{eqnarray} 
H & = &  \omega_0 \sum_{i} \ax _{i} a_{i}  -g
\sum_{i,\sigma} \cx _{i,\sigma} c_{i,\sigma}
 (\ax _{i}+a_{i})  \nonumber \\
& - & \frac{t}{2\sqrt{z}} \sum_{i,j,\sigma} \left 
 (\cx _{i,\sigma} c_{j,\sigma}  + {\rm H.c.} \right ). 
\label{Holstein model}
\end{eqnarray}
At zero temperature, a crossover occurs from a quasi free electron to
a small polaron as the e-ph coupling constant $g$
increases.  This crossover depends strongly on the adiabaticity
parameter $\omega_0/t$ \cite{Piovra,Capone-Grilli}.  If we restrict to
the nearly adiabatic regime ($\omega_0/t < 1$),
the crossover takes place when the polaron energy $E_P= g^2/\omega_0$ is of
the order of the free electron bandwidth, i.e. for
$\lambda=g^2/\omega_0t \sim 1$.  Although these arguments are based on
the ground-state properties alone, it was pointed out in reference
\cite{PRB} that the mechanism of polaron formation can be better
understood by analyzing the whole excitation spectrum, as deduced from
the knowledge of the interacting propagator $G$.  It was shown that
the polaron crossover is characterized by the successive opening of
energy gaps in the low energy part of the spectral density as the e-ph
coupling is increased. At intermediate values of $\lambda$, although
the polaron mass is still close to the free electron value, the
spectral density starts to exhibit typical polaronic features, i.e.
narrow subbands at low energies, together with a broad contribution at
higher energies. The width $W$ of the first subband, which gives a
measure of the coherent (band-like) behaviour of the polaron, is
rapidly reduced as $\lambda$ moves towards the strong coupling limit.
This picture of the polaron crossover has important consequences on
the optical properties.

The calculation of the optical conductivity within the DMFT is
simplified since, for symmetry reasons and due to the locality of vertices, 
no current vertex corrections enter in the Kubo formula
\cite{DINF-OC,vertex}.
Writing the electron spectral function as
\begin{equation}
\label{eq:spectral-function}
\rho_\epsilon(\nu)=- \frac{1}{\pi} Im \frac{1}{\nu + i0^+-\epsilon 
-\Sigma(\nu)},
\end{equation}
then the optical conductivity per particle takes the form
\begin{equation}
\label{sigma-convol}
    \sigma(\omega)= \!\!
    \frac{\zeta\pi}{n\omega} \!\!  \int \!\!  d\epsilon 
    N_\epsilon \phi_\epsilon
    \!\int \!\! d\nu   \left(f_\nu \!- \!f_{\omega+\nu} \right)
    \!\rho_\epsilon(\nu) \rho_\epsilon(\nu+\omega)
\end{equation}
where  $N_\epsilon=(2/\pi)\sqrt{t^2-\epsilon^2}$ is the free DOS of 
the real lattice, $\phi_\epsilon=(t^2-\epsilon^2)/3$ 
is the corresponding current vertex \cite{Freericks} (which ensures a sum rule
for the total spectral weight \cite{Millis}),  
$f$ is the Fermi function, $n$ is the particle 
density and $\zeta$ is a numerical constant 
that we shall drop in the following discussion \cite{spectral-weight}.
Taking the low density limit ($n\rightarrow 0$) at finite
temperatures, following the procedure described in ref. \cite{Mahan}, 
eq. (\ref{sigma-convol}) becomes
\begin{equation}
\label{sigma-compact}
    \sigma(\omega)=
    \frac{\pi}{\omega} (1-e^{-\beta\omega}) \frac{{\cal D}(\omega,\beta)}
    {{\cal N}(\beta)} 
\end{equation}
where we have defined
\begin{eqnarray} \label{caldi}
   {\cal D}(\omega,\beta) &=&
 \int d\epsilon N_\epsilon \phi_\epsilon
 \int d\nu  e^{-\beta(\nu-E_0)} 
          \rho^{(0)}_\epsilon(\nu) \rho^{(0)}_\epsilon(\nu+\omega) \\
  {\cal N}(\beta)&= &\int d\epsilon N_\epsilon
 \int d\nu  e^{-\beta(\nu-E_0)}  \rho^{(0)}_\epsilon(\nu)
 \label{calenne}
\end{eqnarray}
and $\rho^{(0)}$ is the
spectral function at zero density
(these quantities have been rescaled by $e^{\beta E_0}$ to
obtain a finite result in the low temperature limit).  
The integrals in $\epsilon$ can be 
performed analytically, while the remaining integrals in $\nu$ are computed 
numerically, once the self-energy $\Sigma$ is obtained as a continued 
fraction expansion by applying the iteration scheme reported in 
reference \cite{PRB}.

Our treatment applies to a dilute \textit{non-degenerate}
polaron gas, where exchange effects can be neglected.
Turning to real systems, we expect that our results will be 
relevant at low enough density, 
i.e. when the Fermi temperature
is negligible compared to  
the important temperature scales of the problem (e.g. the 
temperature $\hbar \omega_0/k_B$, 
below which phonon quantum effects become important
\cite{Reik}).

\begin{figure}[t]
\vspace{-.5cm}
  \centering\epsfig{file=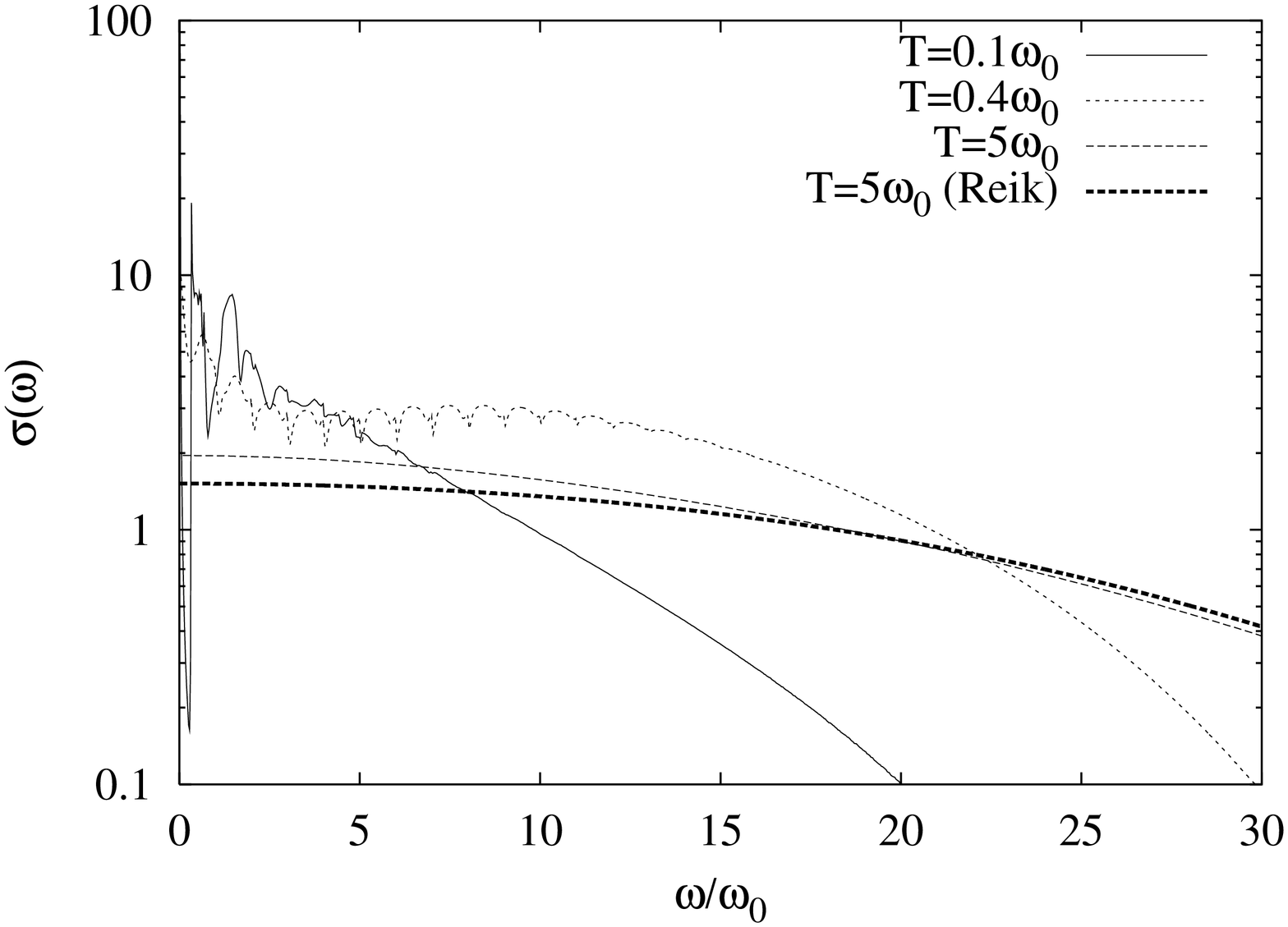,angle=-90,width=0.9\linewidth}
\caption{Optical conductivity per particle
  at $\lambda=0.9$ and $\omega_0/t=0.1$, at different
  temperatures. Reik's result is shown at $T=5\omega_0$ for comparison.
  Note the transfer of spectral weight occurring at low $T$ (for the 
definition of $T_{W}$, see text).
The curve marked $T<<T_{W}$ has been obtained with eq. (\ref{sigma-T0})}
\protect\label{fig:w001l09-hi}
\end{figure}

Figure \ref{fig:w001l09-hi} shows the results obtained 
 at $\omega_0/t=0.1$ and $\lambda=0.9$, i.e.
close to the polaron crossover \cite{note:extreme}.
The optical response shows a typical pattern of subbands roughly
spaced by $\omega_0$, which tends to become smooth at higher
frequencies. 
This behavior is a characteristic signature of the polaron
crossover \cite{PRB}.
At temperatures well above $\omega_0$, the results are in good
agreement with Reik's formula \cite{Reik}, showing that the physics in
this limit  is 
dominated by activated hopping processes.  As the temperature is
decreased,  a substantial transfer of
spectral weight takes place,  the low-frequency features
becoming more  pronounced.

To get some insight on this peculiar behavior, it is convenient to
analyze the polaron excitation spectrum by separating explicitly the
contributions $\rho^>$ and $\rho^<$ at energies respectively above and
below $E_0$.  At low temperatures, $\rho^>$ has quasiparticle poles at
energies below $E_0+\omega_0$ plus a more or less incoherent
background due to phonon scattering at higher energies. 
On the other hand, $\rho^<$ represents
incoherent states where one or more phonons are taken from the thermal
bath.  Correspondingly, by looking at the definitions (\ref{caldi})
and (\ref{calenne}), one can define $ {\cal N}={\cal N}^> + {\cal N}^<$ and
${\cal D}={\cal D}^{>>}+{\cal D}^{<>}+{\cal D}^{<<}$.
The term ${\cal D}^{>>}$ has contributions associated to both coherent
band motion and phonon emission processes, while 
${\cal D}^{<>}$ and ${\cal D}^{<<}$ account for the
thermally activated hopping.

An estimate for the temperature $T_{W}$ below which coherent
processes acquire importance can be obtained by analyzing the relative
weight of ${\cal N}^>$ and ${\cal N}^<$.
Due to the exponential factor in eq. (\ref{calenne}), we see that 
at low temperatures
${\cal N}^>$ weights the coherent states close to $E_0$. 
Using the semicircular free DOS, we obtain ${\cal N}^>=
\sqrt{2(m^*/m)/\pi} \; (T/t)^{3/2}$, where $m^*/m$ is the ratio of the
polaron mass to the free electron value. 
On the other hand, 
being associated to incoherent states, 
the spectral function $\rho_\epsilon^<(\nu)$ 
is essentially $\epsilon$-independent and
peaked around a discrete set of energies $\nu=E_0+W-p\omega_0$, 
with a weight proportional to the thermal occupation factor
$e^{-p\beta\omega_0}$ \cite{footnote-rho} (recall that $W$ is the width
of the first subband in the spectral density, and $p=1,2,..$ is the number 
of thermal phonons).
When evaluating the integral in ${\cal N}^<$, such vanishing
occupation factors are partly compensated by the exponential
$e^{-\beta(\nu-E_0)}$, which gives rise to an overall behavior ${\cal
  N}^< \propto e^{-\beta W}$.

We find that the crossover between the two regimes 
occurs at $T_{W} \sim W/\log(t/W)$. This is not
surprising, since as was noted in the introduction, $W$ is the typical
energy scale for coherent excitations 
(in the particular case studied
here, where $m^*/m\simeq 1.5$ and $W=0.7\omega_0$, we get
$T_{W}\simeq  0.1 \omega_0$).  
For $T\ll T_{W}$, only the terms ${\cal D}^{>>}$ and ${\cal N}^>$
survive in formula (\ref{sigma-compact}), which becomes:
\begin{equation}
  \label{sigma-T0}
     \sigma(\omega)=\pi t \frac{m^*}{m}  \; 
    \left\lbrack\frac{1-e^{-\omega/T}}{\omega/T}\right\rbrack 
    \rho_{q=0}(\omega+E_0)
\end{equation}
where the spectral function $\rho$ is evaluated at $\epsilon=\epsilon(q=0)$.
In this regime, the conductivity  scales as $T/\omega$ 
for any $\omega>T$, 
so that most of the spectral weight is transferred to 
the Drude peak  $\sigma(\omega)\sim 
\delta(\omega)$.

At higher temperatures, the main contribution to the optical absorption
involves phonons that 
are already  present in the thermal bath, so that one expects 
some  spectral weight to move  towards higher frequencies as the
temperature is raised above $T_{W}$. This is indeed the case, as can
be seen in figure 1.  Let us stress that this crossover from
\textit{coherent} to \textit{activated} conductivity is a signature of the
intermediate coupling regime. In
the strong coupling limit, where $W\rightarrow 0$, the crossover
temperature vanishes and phonon-assisted contributions always
dominate. 

\begin{figure}[htbp]
\vspace{-.2cm}
  \centering\epsfig{file=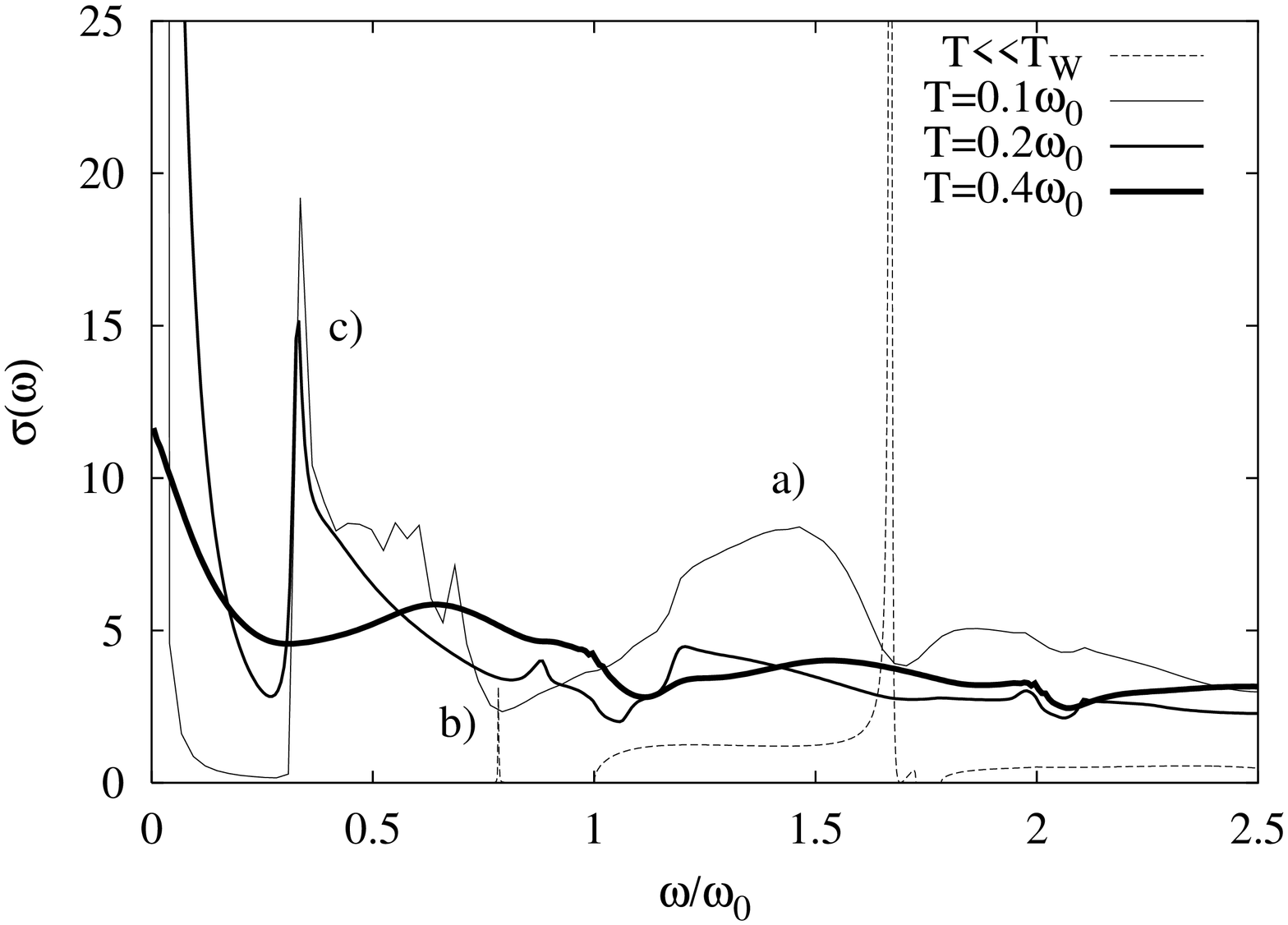,angle=-90,width=0.9\linewidth}
\caption{The same as preceding figure but in the region close to
  $\omega_0$ and with a finer temperature scan.
The curve $T\ll T_{W}$ corresponds to $T=0.025\omega_0$ 
(the Drude peak is not shown).}  
\protect\label{fig:w001l09-lo}
\end{figure}
In Fig.  \ref{fig:w001l09-lo} we have reported the optical
conductivity in the frequency range around $\omega_0$, 
at temperatures close to $T_{W}$.
Despite its apparent complexity, the observed pattern can be
qualitatively understood by looking at the  excitation spectrum close
to the ground state, which is sketched in figure 3 versus the free-particle
energy $\epsilon$ (see eq. (\ref{eq:spectral-function})). 
The dark curves are the quasiparticle states in the lowest order polaron
subbands close to the ground-state, while the  region above $\omega_0$
represents  the
incoherent background due to scattering from Einstein phonons (the
states below $E_0$, which give rise to the hopping-like behavior at
$T\gg T_{W}$, are not shown in fig. 3).  The
transitions from the ground-state to the incoherent background, denoted by (a),
give rise to a continuum of absorption above $\omega_0$.
A transition from the ground-state to the bottom of
the second subband is also possible (b), 
but this is hardly visible in fig. 2 owing to its negligible spectral weight.
In addition, the Drude peak corresponding to coherent band 
motion is clearly visible in fig. 2. 
\begin{figure}[htbp]
\vspace{-.2cm}
  \centering\epsfig{file=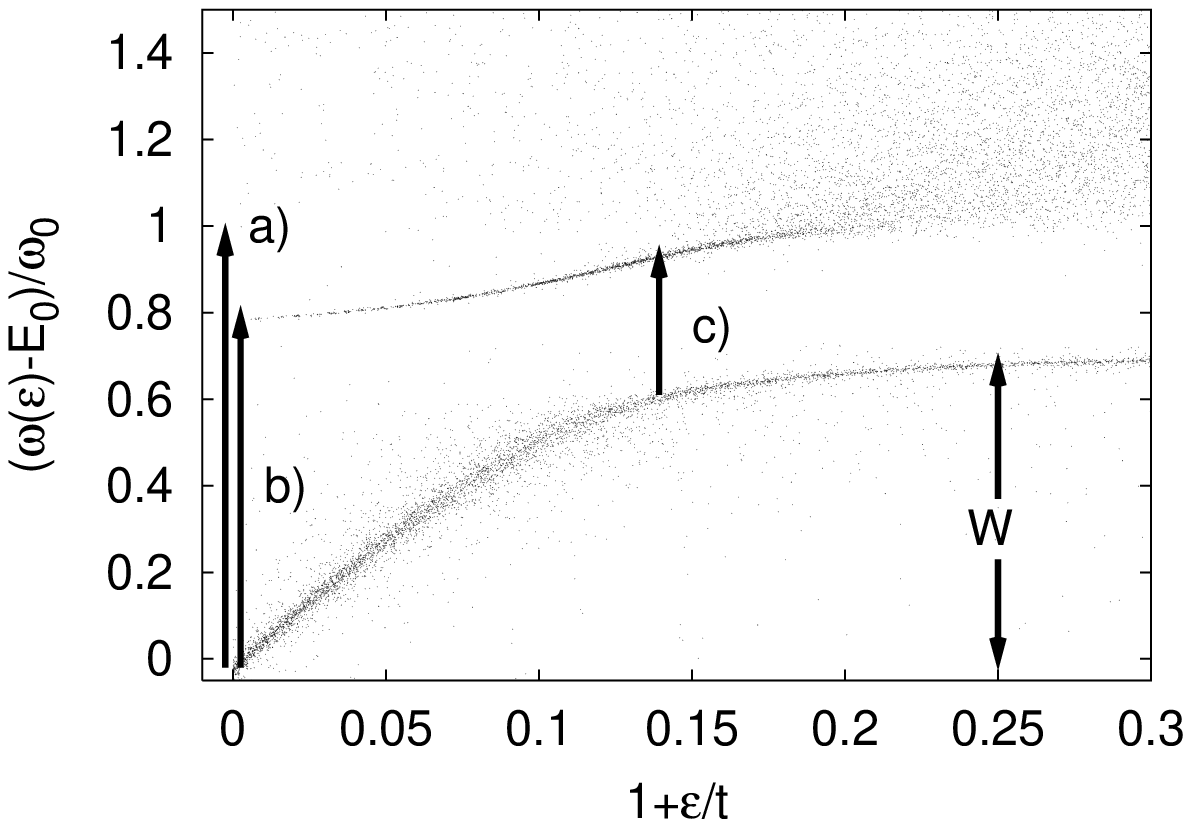,angle=0,width=0.9\linewidth}
\caption{The dispersion of the lowest order polaronic subbands versus
  the free particle energy $\epsilon$ (parameters are the same as
  in the preceding figures, and $T=0.2 \omega_0$). The points are distributed
  according to the spectral function $\rho_\epsilon(\nu)$.}
\protect\label{fig:TAIR}
\end{figure}

The most striking feature in figure 2 is probably 
the appearance of a resonance \textit{below}
$\omega_0$ (c), which is due to \textit{Polaron 
Interband Transitions} (PIT), as can be seen in figure 3.
Indeed,  because of thermal activation,
long-lived states at energy $E_0<\nu<E_0+\omega_0$ can be occupied
with a probability $e^{-\beta(\nu-E_0)}$.
According to the Fermi golden rule, the transition probability at a
given frequency $\omega$ is proportional to the 
joint density of the initial and final
states separated by $\omega$.  If we focus on transitions between
different subbands, this function has a Van Hove singularity
at that value $\omega^*$ where the band dispersions are parallel, which
identifies an \textit{interband threshold} (see arrow (c)).  
This resonance has a non monotonic temperature dependence, which can
be understood as follows: the occupation probability of the initial
state gives rise to an activated behavior at low $T$; 
as the temperature is raised above $\omega^*$, 
the difference in population of the initial and final states is balanced,
which depresses the optical absorption according to the prefactor
$\sim (1-e^{-\beta\omega^*})$ in formula (\ref{sigma-compact}). 
According to the discussion above, we argue 
that PIT are not peculiar to the
Holstein model, nor are they a consequence of the approximations used.
Rather, they are closely related to the presence of several subbands in the
spectral density at low energy, which should be a common feature of
any \textit{finite band} model in the intermediate e-ph coupling
regime. Such interband transitions  
are obviously absent at weak coupling, where there are no
subbands in the excitation spectrum, while in strong coupling they can
hardly be resolved from the incoherent multiphonon absorption,
due to their relatively small spectral weight. On the other hand, the results
are not qualitatively affected by variations of the adiabaticity parameter,
provided that $\omega_0/t<1$ \cite{PRB}.

In this work, we have analyzed the optical conductivity of a dilute
gas of small polarons 
in the framework of the DMFT, focusing on the crossover regime
between quasi-free electrons and  small polarons which takes place 
at intermediate 
values of the e-ph coupling strength. The spectra at high 
temperatures do not differ significantly from the classical 
picture of hopping-like conductivity, which predicts a featureless absorption 
band at finite frequencies. 
However, new interesting results appear at low 
temperatures, where the physics is governed by two different energy scales. At
temperatures below $\omega_0$, phonon \textit{quantum} 
effects become important, and 
the optical spectra start to exhibit narrow peaks at low frequencies,
together with a broad background at higher frequencies. Upon further 
decreasing the temperature,  a noticeable transfer of spectral weight takes 
place towards low frequencies. 
This occurs at a temperature $T_{W}$ which is related to 
$W$, the width of the first polaron subband in the spectral density, 
i.e. the typical energy scale governing \textit{coherent} behaviour. 

Our  results in the low frequency region (see Fig. 2) bear a strong 
resemblance with the far infrared spectra of
both hole- and electron-doped cuprates at low doping levels,
where narrow peaks   
are observed at typical phonon frequencies,
usually called infrared active vibrations (IRAV)  \cite{Kim,Bazhenov}. 
From a theoretical point of view,
if we focus on such low energy features, i.e. much
lower than the typical energies of electronic correlations (e.g. J
in the t-J model), the inclusion of correlation 
effects would only enter through a renormalization of the hopping parameter.
Therefore, we expect that a description in terms of the Holstein model with
renormalized parameters should correctly account for all the observed 
features in the frequency range of interest \cite{parameters}.  
On the other hand, the present single-polaron treatment, 
which neglects polaron-polaron interactions, is 
appropriate at low doping levels, i.e. precisely 
where the IRAV are best resolved. 

In the measured optical spectra, the
IRAV peaks appear as shoulders on both sides  of the main phonon
lines, and they  have been
ascribed to the lowest order transitions inside the polaron  potential-well
\cite{Calvani-Varenna}. We have shown here that 
 in the intermediate e-ph  
coupling regime,  the polaron absorption exhibits narrow
peaks located above and below the phonon 
frequency \cite{many-phonons}, whose $T$ dependence is non monotonic. 
These peaks come from two different processes:
transitions to incoherent scattering states (denoted by (a)), 
and thermally activated Polaron Interband Transitions (denoted by (c)). 
Although indications of a non monotonic behavior of some IRAV 
already exist \cite{Bazhenov,Giura}, a more detailed experimental analysis
of their temperature dependence is 
needed to confirm these ideas.

We wish to thank P. Calvani, P. Giura, J. Lorenzana, 
S. Lupi for valuable discussions, and J.K. Freericks and A.J. Millis 
for pointing out to us the current vertex for the Bethe lattice.

\end{multicols}


\begin{thebibliography}{99}

\bibitem{Holstein} T.Holstein Ann.Phys. {\bf 8}, 325 (1959), {\it
    ibid.} 343 (1959).  

\bibitem{PRB} S. Ciuchi \textit{et al.},
  Phys. Rev. B \textbf{56}, 4494 (1997) 

\bibitem{Kotliar} A.Georges, \textit{et al.}, Rev.Mod.Phys. {\bf 68}, 13
  (1996) and references cited therein.  

\bibitem{Piovra} M. Capone \textit{et al.}, 
 Europhys. Lett. {\bf 42}, 523 (1998); M. Capone \textit{et al.}, 
J. Supercond. to appear (cond-mat/9812195).

\bibitem{Sumi} H. Sumi \textit{et al.} J. Phys. Soc. Japan \textbf{36}, 770 (1974)  

\bibitem{Reik} H.G.Reik in "{\it Polarons in
    Ionic Crystals and Polar Semiconductors}", ed. J.Devreese,
  North-Holland, Amsterdam (1972); D.Emin Adv.Phys  {\bf 24}, 305 (1975)

\bibitem{Footnote2} The hopping amplitude is rescaled as a function of
  the number of nearest neighbors $z$ to yield a finite DOS 
  and a finite kinetic energy per particle (of order $t$)
  when \dinf

\bibitem{Capone-Grilli} M. Capone \textit{et al.} Phys. Rev. B
  {\bf 56}, 4484 (1997) 

\bibitem{DINF-OC} A. Khurana Phys. Rev. Lett. {\bf 64}, 1990 (1990).

\bibitem{vertex} Arguments similar to the ones developed in ref. \cite{DINF-OC}
suggest that vertex corrections, in the case of Holstein interactions, 
should not play a significant
role in any dimensions $d\ge 2$, since in that case 
small polarons are essentially
local entities, giving rise to k-independent vertices. 
This supports the idea that our DMFT results give (at least)
a good qualitative description of finite dimensional systems. 

\bibitem{Freericks} W. Chung and J.K. Freericks, Phys. Rev. B {\bf 56}, 
11955 (1998)
 
\bibitem{Millis} A. Chattopadhyay  \textit{et al.}, Phys. Rev. B {\bf 61}, 10738 (2000)

\bibitem{spectral-weight} M.J. Rozenberg \textit{et al.}, 
Phys. Rev. B {\bf 54}, 8452 (1996)

\bibitem{Mahan} G.D. Mahan, \textit{Many-Particle Physics}, Plenum Press, New
York, 2nd edition (1990)

\bibitem{note:extreme} Although we only focus on the intermediate regime, 
  our results at weak and strong coupling are in
  excellent agreement with the standard expansions. Their presentation is
  left for a future publication.

\bibitem{footnote-rho}
  A few peaks can be seen  below $E_0$ in figure 16 of
  ref. \cite{PRB}.   

\bibitem{Kim} Y.H. Kim \textit{et al.}, Phys. Rev. B \textbf{36}, 7252 (1987);
 P. Calvani  \textit{et al.}, Europhys. Lett. {\bf 31}, 473 (1995)

\bibitem{Bazhenov} A.V. Bazhenov \textit{et al.}, Physica C 214, 45 (1993)

\bibitem{parameters} The choice $\omega_0/t =0.1$ is representative for the
cuprates, where phonons in the Cu-O planes range between $\omega_0=0.01$ 
and $0.06 eV$, 
and the hopping parameter, renormalized by electronic correlations, is 
of the order of $0.1-0.2 eV$, which gives $\omega_0/t=0.05-0.6$.  

\bibitem{Calvani-Varenna} P. Calvani, Proc. of the Int. School of Physics 
``Enrico Fermi'', CXXXVI, eds. G. Iadonisi, J.R. Schrieffer and M.L. Chiofalo, 
IOS Press (Amsterdam 1998)    

\bibitem{many-phonons} In a real material, where different phonon modes are
  coupled to the electrons, a similar pattern should be observed
  in the vicinity of all the phonon frequencies.

\bibitem{Giura} P. Giura, PhD thesis, Orsay (2000) 


\end{thebibliography}
\end{document}